\documentclass[aps,prd,twocolumn,showpacs,groupedaddress,nofootinbib,longbibliography]{revtex4-1}
\pdfoutput=1
\newsavebox{\foobox}

\newcommand{\me}{\mathrm{e}}
\newcommand{\md}{\mathrm{d}}

\usepackage{isomath}
\usepackage{graphicx}
\usepackage{xcolor}
\usepackage{rotating}
\usepackage{stmaryrd}
\usepackage{amsmath,amssymb,graphics}
\usepackage{tensor}

\usepackage[colorlinks=true, urlcolor=violet, linkcolor=blue, citecolor=red, hyperindex=true, linktocpage=true]{hyperref}

\def\ben{\begin{equation}}
\def\een{\end{equation}}

\def\be{\begin{equation}}
\def\ee{\end{equation}}
\def\beq{\begin{equation}}
\def\eeq{\end{equation}}
\def\ba{\begin{array}}
\def\ea{\end{array}}

\def\dalemb#1#2{{\vbox{\hrule height .#2pt
\hbox{\vrule width.#2pt height#1pt \kern#1pt
\vrule width.#2pt}
\hrule height.#2pt}}}

\newcommand{\bea}{\begin{eqnarray}}
\newcommand{\eea}{\end{eqnarray}}
\newcommand{\mt}[1]{\textrm{\tiny #1}}

\newcommand{\rh}{r_\mt{H}}

\begin{document}

\title{Hyperscaling violating black hole solutions and Magneto-thermoelectric DC conductivities in holography}

\author{Xian-Hui Ge}
\affiliation{  
 Department of Physics, Shanghai University, Shanghai 200444, P.R. China}
\affiliation{ Department of Physics, University of California at San Diego, CA92037, USA}

\author{Yu Tian}
\affiliation{School of Physics, University of Chinese Academy of Sciences, Beijing, 100049, P.R.  China}

\author{Shang-Yu Wu}
\affiliation{Department of Electrophysics, National Chiao Tung University, Hsinchu 300, Taiwan, ROC}

\author{Shao-Feng Wu}
\affiliation{ Department of Physics, Shanghai University, Shanghai 200444,  China}
\affiliation{Shanghai Key Lab for Astrophysics, 100 Guilin Road, 200234 Shanghai, P. R. China}

\date{\today}

\begin{abstract}
We  derive new black hole solutions in Einstein-Maxwell-Axion-Dilaton theory with a hyperscaling violation exponent. We then examine the corresponding anomalous transport exhibited by cuprate strange metals in the normal phase of high-temperature superconductors via gauge/gravity duality.
Linear  temperature dependence resistivity and quadratic temperature dependence inverse Hall angle can be achieved. In the high temperature regime, the
heat conductivity and Hall Lorenz ratio are proportional to the temperature. The  Nernst signal first increases as temperature goes up  but it then decreases with increasing temperature in the high temperature regime.
\end{abstract}

\maketitle

\section {Introduction} The  surprising transport behavior of cuprate strange metal has been subjects of intense scrutiny
\cite{brooks}. The normal state transport of high-temperature superconductors is highly anomalous and remains one of the most challenging topics in modern condensed matter physics. The drawbacks of the conventional methods is that based on the perturbation theory, it is hard to deal with strongly coupled many-body problems. The ubiquitous T-linear resistivity, the strong $T^2$-dependence of the Hall angle $\vartheta_H$, the temperature-scaling of magnetoresistance, and thermopower are some of the striking anomalies that have puzzled the  physics community over the past three decades.

For decades, researchers have expended theoretical efforts on determining the relationship between the anomalous transport scaling and the quantum criticality \cite{sachdev,damle,schadev08,philips,marel,keimer}. Recently,  the gauge/gravity duality, as a non-perturbative tool, was suggested to study the Hall angle and dc conductivity in strongly interacting critical theories \cite{blake2014,jianpin,Hartnoll09,McGreevy,Herzog:2009}. Analytic expressions were obtained for thermo-electric conductivity tensors with a magnetic field and disorder \cite{donos,blake2015,kimYK,musso2014,Amoretti2016}, and it was determined that the Hall angle generically displays a different temperature dependence of resistivity \cite{blake2014}. This is in agreement with the two-relaxation-times model suggested by Anderson et al. \cite{anderson97,anderson}. Intensive studies have been conducted for obtaining numerical and analytical expressions for the transport properties of holographic theories with the translational invariance broken \cite{qlattice,lingprl,withers,kim,Gouteraux14,Gouteraux15,GLNS,amoretti,amoretti1,gautlett1,gautlett2,gautlett3,blaise2,blaise3,schalm,gsw2015}.
In \cite{hartnoll, karch14}, a scaling hypothesis was proposed to explain the unusual scaling laws for the thermopower, magnetoresistivity, and Hall Lorenz ratio. Unfortunately, this hypothesis
failed in reproducing the temperature dependence of thermodynamic quantities such as specific heat and entropy density of cuprates. An alternative scaling approach was later proposed for constructing viable phenomenologies of the cuprate strange metal in \cite{Dimi2015,Philip2016}. A more convincing comparison with experimental phenomenologies is still lacking.

In this study, we  employ a new black-hole solution in the IR regime as a toy gravitational dual model and  develop a new computational tool
to study the transport coefficients of the normal state of high-temperature superconductors.
The radial flow for electric conductivity is derived from holographic Wilsonian renormalization group equation. The dc conductivity can be read off from the regularity condition at the event horizon.
As a toy model toward comparisons with real experimental data, we show that some aspects of the temperature-dependence of transport and thermodynamic quantities observed in the experiments can be reproduced in this model. The heat conductivity and Hall Lorenz ratio are proportional to the temperature in the high-temperature regime. The Nernst signal increases as the temperature decreases. Linear temperature resistivity can be obtained at large temperatures and the Fermi-liquid-like $T^2$ resistivity is recovered in the low-temperature regime with weak momentum relaxation. A key observation is that the transport quantities are governed by the same quantum critical dynamics as those for the thermodynamic quantities. For example, entropy density and specific heat are proportional to temperature, which is similar to the proportionality between resistivity and thermal conductivity under the high-temperature limit. In addition to evaluating the temperature dependence of transport, we compute the Lorenz number and Hall Lorenz ratio at zero temperature.

\section{Dyonic black hole solution with hypercaling violating factor}\label{sec:incoh}

In this section, we present the details about how to derive a dyonic black hole solution with dynamic exponent and hyperscaling violating factor in the presence of a magnetic field in $4$-dimensional bulk spacetime.  We consider the
 Einstein-Maxwell-dilaton-axion model with two $U(1)$
 gauge fields. One gauge field coupled with the dilaton
 field is required to generate a Lifshitz-like vacuum, while
 the other has a charged black hole. For further investigation of the transport properties, we  also add linear axions to generate proper momentum relaxation. Therefore, we consider the following action
\bea\label{action}
  S&=&\int\md^{4}x\sqrt{-g}\big[{R}-\frac{1}{2}(\partial\phi)^2
  -\frac{1}{4}\sum_{i=1}^{2}\me^{\lambda_{i}\phi}(F_{i})^{2}\nonumber\\&
  -&\frac{1}{2}\me^{\eta\phi}\sum_{i=1}^{2}(\partial\chi_{i})^{2}
  +\sum_{i=1}^{2}V_{i}\me^{\gamma_{i}\phi}\big],
\eea
where $\lambda_{i},\eta,\gamma_{i},V_{i}$ are undetermined constant parameters and $Z_i(\phi)=\me^{\lambda_{i}\phi}$, $Y(\phi)=\me^{\eta\phi}$ and $V=V_{i}\me^{\gamma_{i}\phi}$.

The equations of motion read
\bea
  &&R\indices{^{\mu}_{\nu}}-\frac{1}{2}\sum_{i=1}^{2}\me^{\lambda_{i}\phi}(F_{i})^{\mu\rho}(F_{i})_{\nu\rho}
  -\frac{1}{2}\me^{\eta\phi}\sum_{i=1}^{2}\partial^{\mu}\chi_{i}\partial_{\nu}\chi_{i} \nonumber \\
  &-&\frac{1}{2}\partial^{\mu}\phi\partial_{\nu}\phi +\frac{1}{2}\delta^{\mu}_{ \nu}\big(\frac{1}{4}\sum_{i=1}^{2}\me^{\lambda_{i}\phi}(F_{i})^{2}
  \nonumber \\ &+&\sum_{i=1}^{2}V_{i}\me^{\gamma_{i}\phi}\big)=0,\label{metricequation}\\
  &&\partial_{\mu}(\sqrt{-g}\me^{\lambda_{i}\phi}(F_{i})^{\mu\nu})=0,\\
  &&\partial_{\mu}(\sqrt{-g}\me^{\eta\phi}g^{\mu\nu}\partial_{\nu}\chi_{i})=0,\\
   &&\frac{1}{\sqrt{-g}}\partial_{\mu}(\sqrt{-g}g^{\mu\nu}\partial_{\nu}\phi)
  -\frac{1}{4}\sum_{i=1}^{2}\lambda_{i}\me^{\lambda_{i}\phi}(F_{i})^2
   \nonumber \\
  &-&\frac{1}{2}\eta\me^{\eta\phi}\sum_{i=1}^{2}(\partial\chi_{i})^2
  +\sum_{i=1}^{2}V_{i}\gamma_{i}\me^{\gamma_{i}\phi}=0.
\eea
At the same time, we assume the following ansatz for the metric, gauge fields and axions
\bea
  \md s^{2}&=&r^{-\theta}\big(-r^{2z}f(r)\,\md t^{2}+\frac{\md r^{2}}{r^{2}f(r)}+r^{2}\,\md x^{2}+r^{2}\,\md y^{2}\big),\nonumber\\
  A_{1}&=&a_{1}(r)\,\md t, \qquad
  A_{2}=a_{2}(r)\,\md t+Bx\,\md y, \nonumber\\
  \chi_{1}&=&\beta x, \qquad
  \chi_{2}=\beta y,\nonumber
\eea
where  $z$ and $\theta$ are dynamical and hyperscaling violation exponents, respectively  \cite{dong,ali,honglu}. The first gauge field $A_{1}$ is an auxiliary gauge field, leading to a Lifshitz-like vacuum.
 The second gauge field is the physical one which provides the finite chemical potential and $B$ is the magnetic field.
From the  equations of motion for gauge fields
\begin{equation}\label{gaugeequation2}
  \partial_{r}(\sqrt{-g}\me^{\lambda_{i}\phi}(F_{i})^{r\nu})=0,
\end{equation}
we obtain the expression for the charge density
\begin{equation}\label{charge}
  q_{i}=J_{i}^{t}=\sqrt{-g}\me^{\lambda_{i}\phi}(F_{i})^{tr}=r^{-z+3}\me^{\frac{1}{2}\lambda_{i}\phi}a_{i}',
\end{equation}
where $q_{i}$ are integral constants. According to holographic principle, $q_{2}$ is interpreted as the charge of the black hole while $q_{1}$ is the ``charge" associated with the auxiliary $U(1)$ gauge field. The ``charge" $q_1$ should be vanishing in the case $z=1$.

Next, subtracting the $rr$ component from the $tt$ component of Einstein equation (\ref{metricequation})
\begin{equation}\label{tt-rr}
  -\frac{1}{2}(\theta-2)(\theta-2z+2)r^{\theta}f+\frac{1}{2}r^{\theta+2}(\phi')^{2}f=0,
\end{equation}
one can solve the dilaton field
\begin{equation}\label{scalarsolution}
  \phi=\sqrt{(\theta-2)(\theta-2z+2)}\ln{r}=\nu\ln{r}.
\end{equation}
The expressions (\ref{charge}) and (\ref{scalarsolution}) change the $xx$ component Einstein equation (\ref{metricequation}) to
\bea
  &&\frac{1}{2}(\theta-2)r^{2\theta-z-1}(r^{-\theta+z+2}f)'-\frac{1}{4}\sum_{i}r^{-\lambda_{i}\nu+2\theta-4}(q_{i})^{2}
  -\nonumber \\
  &&\frac{1}{4}r^{\lambda_{2}\nu+2\theta-4}B^{2} -\frac{1}{2}r^{\eta\nu+\theta-2}\nu^2+\frac{1}{2}\sum_{i}V_{i}r^{\gamma_{i}\nu}=0. \label{xx}
\eea
Solving (\ref{xx}), we obtain the function $f$ in terms of some undetermined parameters
\bea\label{fsolution1}
  f&=&  \sum_{i}\frac{(q_{i})^{2}r^{-\lambda_{i}\nu+\theta-4}}{2(\theta-2)(-\lambda_{i}\nu+z-2)}
  +\frac{B^{2}r^{\lambda_{2}\nu+\theta-4}}{2(\theta-2)(\lambda_{2}\nu+z-2)} \nonumber \\
     & +&\frac{\beta^2r^{\eta\nu-2}}{(\theta-2)(\eta\nu-\theta+z)}
     -\sum_{i}\frac{V_{i}r^{\gamma_{i}\nu-\theta}}{(\theta-2)(\gamma_{i}\nu-2\theta+z+2)}\nonumber \\
     &-& mr^{\theta-z-2},
\eea
where $m$ is an integral constant related to the mass of black hole. The condition that metric is asymptotic to the Lifshitz geometry fixes the parameter $\gamma_{1}$ as follows
\begin{equation}\label{parametersolution1}
  \gamma_{1}=\frac{\theta}{\nu}.
\end{equation}

The determination of remaining parameters need to use the equation of motion of the dilaton field in which the expressions (\ref{charge}) and (\ref{scalarsolution}) are plugged in
\bea
  &&\nu r^{2\theta-z-1}(r^{-\theta+z+2}f)'+\frac{1}{2}\sum_{i}\lambda_{i} r^{-\lambda_{i}\nu+2\theta-4}(q_{i})^{2}
  \nonumber\\ &-&\frac{1}{2}\lambda_{2}r^{\lambda_{2}\nu+2\theta-4}B^{2}
  -\eta r^{\eta\nu+\theta-2}\nu^2\nonumber\\ &+&\sum_{i}V_{i}\gamma_{i}r^{\gamma_{i}\nu}=0. \label{scalarequation2}
\eea
Combining (\ref{xx}) with (\ref{scalarequation2}) by eliminating the function $f$, one can obtain the following relation
\bea\label{parameterequation}
  &&\frac{1}{2}\sum_{i}[\nu+\lambda_{i}(\theta-2)]r^{-\lambda_{i}\nu+2\theta-4}(q_{i})^{2}
  \nonumber\\
  &+&\frac{1}{2}[\nu-\lambda_{2}(\theta-2)]r^{\lambda_{2}\nu+2\theta-4}B^{2} +[\nu-\eta(\theta-2)]r^{\eta\nu+\theta-2}\beta^2
  \nonumber\\
  &+&\sum_{i}V_{i}[\gamma_{i}(\theta-2)-\nu]r^{\gamma_{i}\nu}=0.
\eea
Since $q_{2}$ and $\beta$ could be arbitrary valued constants, their coefficients should be zero and we can determine the values of $\lambda_{2}$ and $\eta$. Although $B$ is also arbitrary, its coefficient has no more undetermined parameter after $\lambda_{2}$ is determined, so we let the coefficients and the exponentials of the terms which contain $B$ and $V_{2}$ be equivalent respectively. Meanwhile the $q_1$ terms should cancel with the $V_1$ term. In the end, we have
\begin{gather}
  \lambda_{2}=\frac{\nu}{2-\theta}, \qquad
  \eta=-\lambda_{2}=\frac{\nu}{\theta-2}, \qquad
  \lambda_{1}=\frac{\theta-4}{\nu},  \nonumber \\
  \gamma_{2}=\frac{\theta+2z-6}{\nu}, \qquad
  q_{1}^{2}=\frac{2V_{1}(z-1)}{z-\theta+1}, \nonumber \\
  V_{2}=\frac{B^{2}(2z-\theta-2)}{4(z-2)}. \label{parametersolution2}
\end{gather}
Since we are considering an effective low temperature physics of the total geometry, it is not surprising that $V_2$ depends on the magnetic fields. The magnetic field is fixed in the IR  and only when $2z-\theta=2$, this term vanishes. For example when $z=1$ and $\theta=0$, $V_2$ disappears.
For simplicity, we would like introduce the transformation
\be
q^2_2\rightarrow 2 (\theta-2)(\theta-z) q^2_2, ~~~~~~~\beta^2 \rightarrow (\theta-2)(z-2)\beta^2.
\ee
Then, we can plug all parameters (\ref{parametersolution1}) and (\ref{parametersolution2}) into the expression (\ref{fsolution1}) and obtain the blacken factor function
\bea\label{fsolution2}
  f&=&1-mr^{\theta-z-2}+{q^2_{2}r^{2\theta-2z-2}}+\frac{B^{2}r^{2z-6}}{4(z-2)(3z-\theta-4)}\nonumber \\
  &-&{\beta^2r^{\theta-2z}}.\nonumber
\eea
The constant term is set to be one, as long as we demand
\begin{equation}\label{v1solution}
  V_{1}=(z-\theta+1)(z-\theta+2).
\end{equation}



One may notice that as  $(z-\theta)\rightarrow 0$, $f(r)$ appear to diverge.   A  well-defined solution can be achieved in an alternative form:
\bea
f(r)&=&1-{\frac{m}{r^{2+z-\theta}}}-\frac{q^2_2\ln r}{2(2-\theta)r^{d+z-\theta}}-{\frac{ \beta^2}{r^{2z-\theta}}}\nonumber\\ &+&\frac{B^{2}r^{2z-6}}{4(z-2)(3z-\theta-4)},\label{flog}\\
&=&1-{\frac{m}{r^{2}}}-\frac{q^2_2\ln r}{2(2-z)r^{2}}-{\frac{ \beta^2}{r^{2z-\theta}}}\nonumber\\ &+&\frac{B^{2}r^{2z-6}}{4(z-2)(3z-\theta-4)},\nonumber\\
F_{2rt}&=&q_2 r^{-1}, \label{frt}
\eea
where $m$ and $q_2$ are finite physical parameters without divergence as $(z-\theta)\rightarrow 0$. A careful examination of (\ref{flog}) and (\ref{frt}) reveals that they satisfy the corresponding Einstein equation and Maxwell equation.
The Hawking temperature at the event horizon $r=\rh$ is given by
\bea
T&=&\frac{\rh^z}{2\pi}\bigg(1-\frac{q^2_2}{4(2-z)\rh^2}-\frac{\beta^2 (2+\theta-2z)}{2 \rh^{2 z- \theta }}\nonumber\\ &+&\frac{B^{2}\rh^{3z-7}}{4(z-2)(3z-\theta-4)}\bigg).
\eea
As $z \rightarrow 1$ and $\theta \rightarrow 1$, the first auxiliary gauge field then vanishes and the metric of the black hole solution can be recast as
\bea\label{metric}
ds^2&=&-g_{tt}dt^2+g_{rr}dr^2+g_{xx}(dx^2+dy^2)\nonumber\\&=&\!r^{-\theta}\bigg[-r^2f(r)dt^2+\frac{dr^2}{r^2f(r)}+r^2(dx^2+dy^2)\bigg]\!,\nonumber\\
f(r)&=&1-\frac{m}{r^2}-\frac{q^2_2 \ln r}{2 r^2}-\frac{\beta^2}{r}+\frac{B^2}{8r^4},\\
 A=&q_2& \ln r dt+\frac{B}{2}(xdy-ydx), e^{\phi}=r,  \chi_i=\beta \delta_{ia} x^a, \nonumber\\ V(\phi)&=&2r+B^2 r^{-3}/8, Z(\phi)=r^{-1}, Y(\phi)=r, \nonumber
\eea
where $\theta$, $m$, $q$, and $B$ are the parameters related to the hyperscaling violation factor, mass, charge, and magnetic field, respectively. (\ref{metric}) is the main metric ansatz used in this paper.
Note that the magnetic field $B$ appears in the potential $V(\phi)$ because we only consider the IR geometry here so that the magnetic field  is fixed in the action.
 One critical observation is that in the absence of $\beta$ and $B$ terms, the blacken function $f(r)$ and the scalar potential $A_t$ are the same as those of the
 $(2+1)$-dimensional charged BTZ black brane \cite{btz,lut,seng}. The electrostatic potential of the black hole diverges asymptotically as $\ln r$. But the presence of divergent
 boundary terms is an artifact of the remormalization procedure and the divergence can be removed \cite{seng2}. The boundary is located at $r\rightarrow +\infty$. The nondegenerate horizon is located at $r=\rh$, where $f(\rh)=0$ and its associated Hawking temperature is given by
\be
T=\frac{\rh}{2\pi}\bigg(1-\frac{q^2_2}{4\rh^2}-\frac{\beta^2}{2\rh}-\frac{B^2}{8\rh^4}\bigg).
\ee
The black hole solution is not an asymptotic AdS solution and therefore can in principle be interpreted as an IR geometry  embedded in the AdS space.
When $\rh\gg \beta^2$, $\rh^2 \gg q^2_2$ and $\rh^4 \gg B^2$, we recover the well known relativistic scaling $T\sim \rh$. This corresponds to large temperature regime although the temperature satisfying these ranges can be decreased by tuning $\beta$, $q_2$ and $B$.
The entropy density and the specific heat of this black hole are given by $s=4 \pi \rh$ and \be
c_{q_2,\beta,B}=T \bigg(\frac{\partial s}{\partial T}\bigg)_{q_2,\beta,B}=4\pi ^2 \left(1+\frac{\beta ^2+4 \pi  T}{\sqrt{4 q^2_2+\left(\beta ^2+4 \pi  T\right)^2}}\right).\nonumber
 \ee These are proportional to the temperature in the ``large temperature" regime. Experiments on optimally doped  $\rm Y\!Ba_2 CuO_{7-\delta}$ provide the electronic specific heat $c \sim T$ from critical to room temperatures \cite{loram1994}. As shown in the following, this Fermi-liquid scaling does not conflict with that of the anomalous transport.

 The electric charge density is $q_2\equiv -J^t=-\sqrt{-g}Z(\phi)\partial_r A_t$. At zero temperature, that is, when $T=0$, the solution near the horizon develops an $AdS_2 \times R^2$ geometry. The near-horizon geometry reads $ds^2\sim u^{-2}(-dt^2+du^2)+\rh^2 (dx^2+dy^2)$, where $u=\rh^2/(r-\rh)$. The Hawking temperature shows that even in the absence of the $U(1)$ gauge field, the black hole could still be extremal with a near-horizon geometry of $AdS_2$. This implies that at low temperatures, the theory flows to an infrared fixed point in the presence of linear axion fields.
 However, a near extremal dyonic black hole is unstable to forming neutral scalar hair. From the holographic dictionary point of view, we expect there is a domain wall solution interpolating between $AdS_4$ in the UV and
dyonic $AdS_2 \times {R}^2$ solutions in the IR. In \cite{gauntlett}, dyonic black holes at finite temperature approaching $AdS_4$ in the UV were constructed. If the scalar field is above the BF bound for $AdS_4$ but below the  Breitenlohner-Freedman bound for $AdS_2$, the black hole becomes unstable near extremality. For big enough magnetism, the black hole approaches hyperscaling violating behavior in the IR as $T\rightarrow 0$ \cite{gauntlett}.

\section{Holographic transport coefficients}
The transport coefficients are computed in the holography by studying perturbations of the background solution. We  developed the holographic Wilsonian approach previously given in \cite{liu,po,sin,strominger,ge} and computed the dc conductivity by applying linear sources to the boundary fields. As far as we know, the Wilsonian renormalization group approach
 has not been utilized to describe the radial flow for transport coefficients in the presence of momentum dissipation.
 For computing the transport coefficients, the consistent perturbation ansatz reads as
\bea
&& \delta A_{x_i}=\int^{\infty}_{-\infty}\frac{d \omega}{2\pi}e^{-i\omega t}a_i(\omega,r),\\
&& \delta g_{t x_i}=\int^{\infty}_{-\infty}\frac{d \omega}{2\pi}e^{-i\omega t}g_{xx}h_{t x_i}(\omega,r),\\
&& \delta \chi_{i}=\int^{\infty}_{-\infty}\frac{d \omega}{2\pi}e^{-i\omega t}\chi_{i} (\omega,r).
\eea
where $i$ runs over (1,2), and $x_1=x$ and $x_2=y$. We notice that in the computation of conductivity, we also need to perturb the potential but the $B^2$ term does not contribute to the equations of motion as we can see $\delta A_{x_i}$ is only a function of $r$ and $t$.
Since  the spatial $SO(2)$ is unbroken, it is convenient to organize the fields as
\bea
\! a_z=\frac{(a_x-i a_y)}{2}, h_{tz}=\frac{(h_{tx}-i h_{ty})}{2},\chi_z=\frac{(\chi_x-i \chi_y)}{2}.\!\nonumber
\eea
We can define a matrix $\tilde{\sigma}$ from
\[
\left \llbracket \begin{array}{c}
\sqrt{g_{tt}/g_{rr}} Z a_{z}^{\prime}\\
g^2_{xx}/\sqrt{g_{rr}g_{tt}}h_{tz}^{\prime}\\
\sqrt{g_{tt}/g_{rr}} Z g_{xx}{\chi}_z^{\prime}
\end{array} \right \rrbracket=\tilde{\sigma}\left \llbracket\begin{array}{c}
i\omega a_{z}\\
i\omega h_{tz}\\
i\omega {\chi}_z
\end{array}\right \rrbracket,
\]
where the special notation $\llbracket ...\rrbracket$ is what we introduced just for convenience, for example
\bea\left \llbracket \begin{array}{c}
 a_{z}\\
 h_{tz}\\
 {\chi}_z
\end{array}\right \rrbracket\equiv
 \left(\begin{array}{ccc}
 a_{z} & a^{(2)}_{z} & a^{(3)}_{z}\\
h_{tz} & h^{(2)}_{tz} & h^{(3)}_{tz}\\
 {\chi}_{z} &{\chi}^{(2)}_{z} & {\chi}^{(3)}_{z}
\end{array}\right),
\eea
where $a^{(i)}_{z}$, $h^{(i)}_{tz}$ and ${\chi}^{(i)}_{z}$ are linearly independent source vectors, introduced to guarantee the source term invertible.
The RG flow equation of $\tilde{\sigma}$ can be obtained by taking derivative of $\tilde{\sigma}$ {(see the Appendix B for details)}\cite{tian}:
\begin{eqnarray*}
\tilde{\sigma}^{\prime}
 & = & \frac{Z\sqrt{g_{rr}/g_{tt}}}{i\omega}\left(\begin{array}{ccc}
-\omega^2 & \omega B&0\\
-\omega B & g_{xx}Y\beta^2+B^2& i \omega g_{xx}Y\beta^2\\
0 & i \omega g_{xx}Y(\phi)\beta^2& -\omega^2 g_{xx}
\end{array}\right)\nonumber\\&-&
i\omega \sqrt{\frac{g_{rr}}{g_{tt}}}\tilde{\sigma}\left(\begin{array}{ccc}
1/Z& 0 & 0\\
0 & g_{tt}/g^2_{xx}&0\\
0 & 0 & \frac{1}{g_{xx}Z}
\end{array}\right)\tilde{\sigma}\nonumber\\&+&\left(\begin{array}{ccc}
0& \frac{\sqrt{g_{rr}g_{tt}}}{g_{xx}} & 0\\
\frac{q_2}{Z}\frac{g_{rr}}{g_{tt}} & g_{tt}/g^2_{xx}&0\\
0 & 0 & \frac{1}{g_{xx}Z}
\end{array}\right)\tilde{\sigma}.
\end{eqnarray*}
The advantage of the holographic Wilsonian approach is that it reduce the computation of conductivity from second order ordinary differential equations to  first order non-linear ordinary differential equations.
The regularity condition at the event horizon gives boundary condition
\[
 \tilde{\sigma}_0=\left(\begin{array}{ccc}
Z & -\frac{ZB}{\omega}&0\\
\frac{BZ-iq_2}{\omega} & \sigma_{22} &-\frac{i  \beta g_{xx} Y}{\omega}\\
-\frac{i \beta g_{xx}Y}{\omega} & 0& g_{xx}Y
\end{array}\right).
\]
From the definition of the matrix $\tilde{\sigma}$, we obtain the boundary condition at the event horizon
\be
\sqrt{\frac{g_{tt}}{g_{rr}}}a'_z \rightarrow i \omega a_z-i B h_{tz}\bigg|_{r=\rh}.
\ee
Finally, we obtain the expressions for $h_{tx}$ and $h_{ty}$ at the event horizon
\bea
h_{tx}&=&-i\omega\frac{a_x g_{xx}q_2 \beta^2 Y+a_y B(q_2^2+B^2 Z^2+g_{xx}Z \beta^2 Y)}{(B^2 Z+g_{xx} \beta^2 Y)^2+B^2 q_2^2}\nonumber\\
&-&i\omega\frac{Bg_{xx}Y\chi_y+(B^2 g_{xx}YZ+g^2_{xx}Y\beta^2)\chi_x}{(B^2 Z+g_{xx} \beta^2 Y)^2+B^2 q_2^2}\bigg|_{r=\rh},\nonumber\\
h_{ty}&=&i\omega\frac{a_x B(q_2^2+B^2 Z^2+g_{xx}Z \beta^2 Y)-a_y g_{xx}q_2 \beta^2 Y}{(B^2 Z+g_{xx} \beta^2 Y)^2+B^2 q_2^2}\nonumber\\
&+&i\omega\frac{Bg_{xx}Y\chi_x-(B^2 g_{xx}YZ+g^2_{xx}Y\beta^2)\chi_y}{(B^2 Z+g_{xx} \beta^2 Y)^2+B^2 q_2^2}\bigg|_{r=\rh}.\nonumber
\eea
The radially conserved currents can be deduced from the Maxwell equation
\bea
&&J_{x}=i\omega Z a_x-q_2 h_{tx}-Z B h_{ty},\nonumber\\
&&J_{y}=i\omega Z a_y-q_2 h_{ty}+Z B h_{tx}.
\eea
An important step in the evaluation of thermoelectric and heat conductivity tensors involves the determination of physical quantities that are independent of the radial coordinate. The conductivity tensor can be evaluated using
$\sigma_{ij}=\partial J_i/\partial E_j$ with $E_j=-i \omega a_j$. The electrical conductivity tensor reads as
\bea
&& \sigma_{xx}=\sigma_{yy}=\frac{\beta^2\rh^2(B^2+q_2^2 \rh^2+\beta^2 \rh^3)}{B^4+\beta^4 \rh^6+B^2 \rh^2(q_2^2+2\beta^2\rh)},\\
&&\sigma_{xy}=-\sigma_{yx}=\frac{B q_2(B^2+q_2^2 \rh^2+2\beta^2 \rh^3)}{B^4+\beta^4 \rh^6+B^2 \rh^2(q_2^2+2\beta^2\rh)}.
\eea
Notably, in the absence of the external magnetic field, the DC electric conductivity along the $x-$direction is separated into two terms
\be
\sigma_{xx}\sim \frac{1}{2\pi T}+\frac{q_2^2}{4\pi^2\beta^2 T^2}.
\ee
The corresponding resistivity in the small $\beta$ limit can be written as
\be
\rho_{xx}\simeq \frac{4\beta^2\pi^2 T^2}{q^2_2+2\beta^2\pi T}=\frac{\tilde{T}^2}{\tilde{T}+\Delta},
\ee
where we defined $\tilde{T}=2\pi T$ and $\Delta=q^2_2/\beta^2$.
For $\tilde{T}\gg \Delta$, the corresponding resistivity is dominated by the linear-T behavior, whereas in the low-temperature regime $\tilde{T}\ll \Delta$, the system is observed to support the Landau's Fermi-liquid $T^2$ law.  On the other hand, in the strong external magnetic field limit (i.e. $B\gg q_2, \beta$) and  in the small temperature limit, the electric resistivity can be approximated as
\bea\label{resis}
\rho_{xx}&=&\frac{\beta^2\rh^2[B^2 q_2^2+\rh^2(q_2^2+\beta^2 \rh)]}{B^2+\rh^2(q_2^2+\beta^2 \rh)^2}\sim \frac{\beta^2}{q_2^2}\rh^2\\
&\sim& 0.35 B \frac{\beta^2}{q_2^2}+1.86T\sqrt{B}\frac{\beta^2}{q_2^2}+2.46T^2 \frac{\beta^2}{q_2^2}+\mathcal{O}(T^3).\nonumber
\eea
 Since the quadratic temperature dependence of inverse Hall angles has been observed in multiple cuprates in the underdoped to overdoped regions \cite{brooks}, the calculation of  Hall angle in holography is also an interesting topic.
 The Hall angle
$\cot \vartheta_H\equiv \sigma_{xx}/\sigma_{xy}$,
 in this model is found to be
\be
\cot \vartheta_H \sim \frac{\beta^2 \rh^2}{B q_2}\sim T^2.
\ee
 Thus we give an explicit model realizing the anomalous scaling of the Hall angle proposed in Ref.\cite{blake2014}. That is to say, the Hall relaxation rate is different from the transport scattering rate of the linear in temperature resistivity.  Below, we also calculate other transport coefficients such as thermo-electric conductivities, thermal conductivity and Lorenz constants.

The conserved heat currents $\mathcal{Q}^i$ are defined by introducing a two-form associated with the Killing vector field equation $K=\partial_t$, as follows \cite{blake2015,kimYK}
\bea
 \mathcal{Q}^i&=&\sqrt{\frac{g_{tt}}{g_{rr}}} \bigg(-\frac{g'_{tt}}{g_{tt}}g_{ii}\delta h_{ti}+g'_{ii}\delta h_{ti}+g_{ii} \delta h'_{tx}\bigg)-A_t J^i
 \nonumber\\ &+&M(r)\epsilon_{ij}E_j+2 M_{Q}(r)\epsilon_{ij}\zeta_j,
\eea
where \be
M(r)=\int^r_{\rh} dx g^{xx}Z(\phi)B,
\ee and \be M_{Q}(r)=\int^r_{\rh} dr' g^{xx}Z(\phi)B A_t(r').\ee
Note that $M(r)$ and $M_Q (r)$ correspond to the magnetization density of the boundary theory as $r\rightarrow\infty$. The aforementioned ansatz corresponds to the application of an external electric field $E_i$ along the $x_i$-direction and the temperature gradient $(\nabla T)_i=\zeta_i T$ to the boundary theory.
As the heat currents $\mathcal{Q}^i$ are radially conserved, we can evaluate them at the event horizon
\bea
&&\mathcal{Q}^x=-\sqrt{\frac{g_{tt}}{g_{rr}}}\frac{g'_{tt}}{g_{tt}}g_{xx}\delta h_{tx}\bigg|_{r=\rh},\nonumber\\
&&\mathcal{Q}^y=-\sqrt{\frac{g_{tt}}{g_{rr}}}\frac{g'_{tt}}{g_{tt}}g_{xx}\delta h_{ty}\bigg|_{r=\rh}.
\eea
The electrothermal conductivity can be evaluated at the event horizon as
\bea \label{thermoelectric}
&&\alpha_{xx}=\alpha_{yy}=\frac{\partial \mathcal{Q}^x}{T \partial E_x}=\frac{4\pi q_2 \beta^2\rh^5}{B^2 q_2^2\rh^2+(B^2+\beta^2 \rh^3)^2},\\
&&\alpha_{xy}=-\alpha_{yx}=\frac{\partial \mathcal{Q}^y}{T \partial E_x}=\frac{4\pi B \rh(q_2^2\rh^2+B^2+\rh^3 \beta^2)}{B^2 q_2^2\rh^2+(B^2+\beta^2 \rh^3)^2}.\nonumber
\eea Finally, we can extract the heat conductivity tensor as
\bea
&&\bar{\kappa}_{xx}=\bar{\kappa}_{yy}=\frac{16\pi^2 T \rh^3(B^2+\beta^2 \rh^3)}{B^2 q_2^2\rh^2+(B^2+\beta^2 \rh^3)^2},\\
&&\bar{\kappa}_{xy}=-\bar{\kappa}_{yx}=\frac{16\pi^2 T B q_2\rh^4}{B^2 q_2^2\rh^2+(B^2+\beta^2 \rh^3)^2}.
\eea
 The usual thermal conductivity
that is more readily measurable experimentally is defined
by introducing the thermal conductivity at zero electric current, $\kappa_{xx}=\bar{\kappa}_{xx}-\alpha_{xx} \bar{\alpha}_{xx}T/\sigma_{xx}$,
\be
\kappa_{xx}=\frac{16 \pi^2 \rh^3 T}{B^2+\rh^2(q_2^2+\beta^2 \rh)}.
\ee
In the large-temperature regime, the heat conductivity ${\kappa}_{xx} \sim T$ which is same as the specific heat.

Next, we evaluate the Hall Lorenz ratio as follows
\be\label{halllorenz}
\bar{L}_H\equiv \frac{\bar{\kappa}_{xy}}{T \sigma_{xy}}=\frac{16\pi^2 \rh^4}{B^2+q_2^2\rh^2+\beta^2\rh^3}.
\ee
At high temperature, the Hall Lorenz ratio behaves as $\bar{L}_H\sim T$. This temperature dependence can be compared with the experimental results in \cite{zhang2000}. At zero temperature, with a vanishing magnetic field, the Hall Lorenz ratio becomes $\bar{L}_{H}=4\pi^2+\frac{4\pi^2\beta^2}{\sqrt{4q_2^2+\beta^4}}$.

Since the Nernst signal in cuprates has significantly different behaviors compared with conventional metals, we examine the Nernst signal holographically
\bea
e_{N}&\equiv& (\sigma^{-1}\cdot \alpha)_{xy}=\frac{4\pi B \rh^3\beta^2}{B^2 q_2^2+\rh^2(q_2^2+\rh \beta^2)^2}
\nonumber\\ &\sim& \frac{4 \pi}{\beta^2 T}+\mathcal{O}(\frac{1}{T^2}),
\eea
where $\sigma$ and $\alpha$ denote the electric and thermoelectric conductivity matrices, respectively.
\begin{figure}
\center{
\includegraphics[scale=0.45]{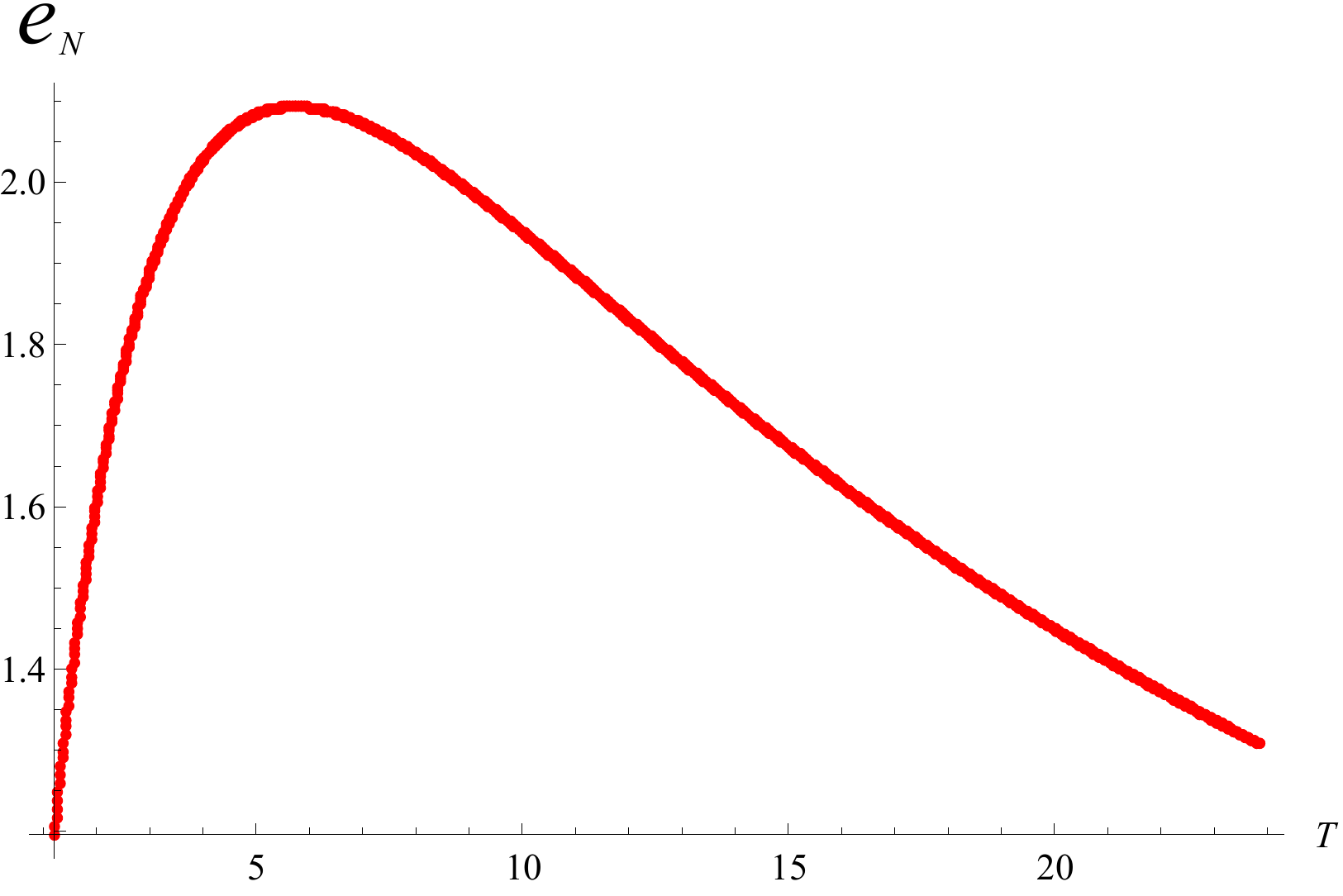}
\caption{\label{solu}  Nernst signal as a function of temperature with $q_2=3$, $\beta=1/2$, and $B=5$.  } }
\end{figure}
\begin{figure}
\center{
\includegraphics[scale=0.45]{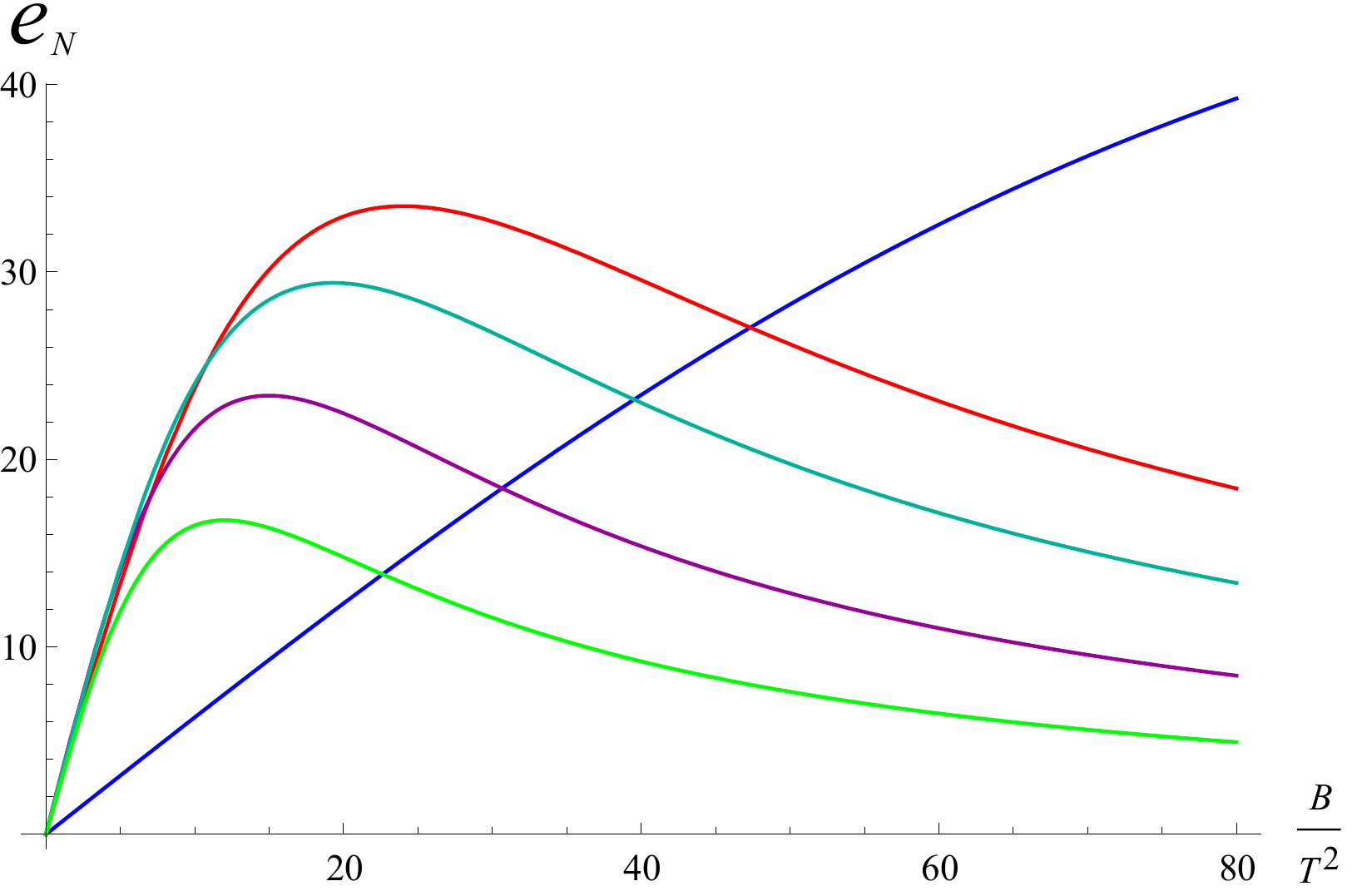}
\caption{\label{solu} Nernst signal as a function of magnetic field B. The lines from the top to bottom correspond to
$\beta/T=1,0.84,0.66,0.5,1.5$. } }
\end{figure}
Figure 1 shows that the Nernst signal increases as temperature decreases at the high temperature limit. Moreover, as shown in Fig. 2, the obtained Nernst
signal demonstrates a bell-shaped dependence on the magnetic field. The blue line is almost straight, whereas the green and red lines yield hill profiles. Our system shows the transition from typical metal (blue line) to the cuprate
state (green, purple and red), as the disorder $\beta$ decreases in the Nernst signal perspective.
  Comparisons can be made between the results obtained here and experimental data presented in \cite{wang, convention,cooper}. For example, the ``hill" profile shown in Fig. 1 and Fig. 2 is similar to those in Figs.5, 6, 9, and 12 in \cite{wang}.  But it is not very clear about the scaling behavior of the Nernst signal from the existing literature.



Moreover, we are able to consider the thermoelectric power (Seebeck coefficient)£º
\be
S\equiv \frac{\alpha_{xx}}{\sigma_{xx}}=\frac{4\pi q_2 \rh^3}{B^2+\rh^2(q_2^2+\beta^2 \rh)}.
\ee
Under the weak magnetic field and charge limit, $S$ behaves as a constant. But in the small disorder limit but keeping $q_2$ and $B$ fixed, we have $S\sim T$ as $\rh$ becomes large. The experimental result obtained at optimal doping suggested the relation $S\sim a -b T$, where $a$, $b$ are constants\cite{brooks}. However, there is no consensus on the temperature scaling of the thermoelectric power from the experimental data\cite{kimjs}, even though some data appears to show a negative slope.



The Lorenz constant $L$, which is related to the Wiedemann-Franz law is a key signature of a Fermi-liquid. In general, if $L/L_0=1$ ($L_0=\pi^2/3\times k^2_B/e^2$) at zero temperature, Landau's Fermi-liquid description is satisfied.
In \cite{crossno}, $L/L0 \gg 1$ has been used to diagnose strong deviations from the quasiparticle picture in graphene.
On the other hand, $L/L_0<1$ at zero temperature also indicates the breakdown of the Fermi-liquid theory \cite{senthil,pfau,lucas}.

At zero temperature and under a vanishing magnetic field, the Lorenz ratio  is given by
\be
L=\frac{\kappa_{xx}}{T \sigma_{xx}}\bigg|_{T,B\rightarrow 0}=4\pi^2 \frac{\beta^4}{4q_2^2+\beta^4}+\frac{4\pi^2 \beta^2}{\sqrt{4q_2^2+\beta^4}}.
\ee
Therefore, at $B=0$, the Lorenz number becomes $8 \pi^2$ only when the disorder becomes stronger (i.e. $\beta^4/q_2^2\rightarrow \infty$).  As $\beta^4/q_2^2 \rightarrow 0$, the Lorenz number
approaches zero and deviations from the Fermi-liquid behavior can be observed.
A Lorenz ratio at zero temperature comparable with the Hall Lorenz ratio given in (\ref{halllorenz})
\be
\bar{L}= \frac{\bar{\kappa}_{xx}}{T\sigma_{xx}}\bigg|_{T,B\rightarrow 0}=4\pi^2+\frac{4\pi^2\beta^2}{\sqrt{4q_2^2+\beta^4}}.
\ee
The Hall Lorenz ratio $\bar{L}_H$ at zero temperature under the weak magnetic field with finite $q_2$ and $\beta$ limits exactly equals to $\bar{L}$ obtained in this study.
 In the large temperature regime, the Lorenz ratios $\bar{L}$ and $L$ presented here are also proportional to the temperature same as $\bar{L}_{H}$.

\section{ Discussions  and conclusions}
In summary, we started from the general Einstein-Maxwell-axion-dilaton theory, an effective low temperature theory, in 4-dimensional spacetime and obtained a dyonic black hole solutions with non-zero hyperscaling violating factor. The temperature-dependence of thermodynamic quantities such as entropy density and specific heat  were  obtained in the small $\beta$, $q_2$ and $B$ limit. Since the metric represents the low temperature part of the geometry and can be connected to a UV $AdS_4$ geometry, the thermodynamic quantities expanded in the high temperature make sense in the calculations.

We then computed the  transport  coefficients in this holographic model.  Linear  temperature dependence resistivity and quadratic temperature dependence inverse Hall angle were achieved.
Temperature scaling of the Hall Lorenz ratio and Nernst signal are also calculated in this model.  The dc transport quantities obtained here are governed by the same quantum critical dynamics as those for the thermodynamic quantities of the black hole. These results can be  compared to that of cuprate strange metals. It seems that, by exploring a new black-hole solution with dynamical exponent $z=1$ and hyperscaling violation exponent $\theta=1$, one can reproduce different temperature scalings of the anomalous
transport observed in experiments.  However, the results  were  obtained by introducing an extra scale related to $\beta^2$, related to the scaling dimension of the operator breaking translations. More precisely, a particular combination of  $\beta^2$, $q_2$ $B$ and $T$ has lead  to  (\ref{resis}). So do the heat conductivity, the Hall Lorenz ratio and the Nernst signal. Therefore, the results cannot be directly  compared to \cite{hartnoll} and \cite{Dimi2015}.  In the future, it would be interesting to compute numerical optical conductivities and the onset of superconductivity in this holographic model.



\textit{Acknowledgements.---} We would like to thank the anonymous referees for valuable comments and suggestions.
The authors also would like to thank Glenn Barnich, John Mcgreevy, Sang-Jin Sin and  Elias Kiritsis for helpful discussions.
The study was partially supported by NSFC,
China (grant No.11375110); NSFC (grant No. 11475179)  and No. 14DZ2260700 from Shanghai Key Laboratory of High Temperature Superconductors; the Ministry of Science and Technology (grant No. MOST 104-2811-M-009-068) and National Center for Theoretical Sciences in Taiwan; and by NSFC
China (grant No.11275120).\\

\appendix

\section{A new analytic method on transport coefficients}
\label{AppA}
It was proved in \cite{sin} that several approaches to RG flow of transport coefficients are equivalent: They are sliding membrane paradigm \cite{liu}, Wilsonian
fluid/gravity \cite{strominger} and Holographic Wilsonian RG \cite{po}.
The essential idea of holographic Wilsonian Renormalization group approach is to integrate out the bulk field from the boundary up to some intermediate radial distance.
The radial direction in the bulk marks the energy scale of the boundary theory and the radial flow in the bulk geometry can be interpreted as the renormalization group flow of
the boundary theory.

The equations of motion of the linear perturbation are given by
\bea
&&\bigg(\frac{g^2_{xx}}{\sqrt{g_{rr}g_{tt}}}h'_{tz}\bigg)'-q_2 a'_z-\beta^2 g_{xx} Y\sqrt{\frac{g_{rr}}{g_{tt}}}h_{tz}
\nonumber\\&&-i\omega g_{xx}Y \sqrt{\frac{g_{rr}}{g_{tt}}}\chi_z+\omega B Z\sqrt{\frac{g_{rr}}{g_{tt}}} a_z-B^2 Z \sqrt{\frac{g_{rr}}{g_{tt}}} h_{tz}=0,\nonumber\\
&& \bigg(\sqrt{\frac{g_{tt}}{g_{rr}}}Z a'_z\bigg)'-q_2 h'_{tz}+\omega^2 Z \sqrt{\frac{g_{rr}}{g_{tt}}} a_z-\omega B Z \sqrt{\frac{g_{rr}}{g_{tt}}}h_{tz}=0,\nonumber\\
&& \bigg(\sqrt{\frac{g_{tt}}{g_{rr}}}g_{xx}Y \chi'_z\bigg)'+\omega^2 Y g_{xx}\sqrt{\frac{g_{rr}}{g_{tt}}}\chi_z-i\omega \beta^2 Y g_{xx} \sqrt{\frac{g_{rr}}{g_{tt}}} \nonumber\\&&h_{tz}=0,\\
&& \frac{\omega g^2_{xx}}{\sqrt{g_{rr}g_{tt}}}h'_{tz}-B\sqrt{\frac{g_{tt}}{g_{rr}}}a'_z+i\beta g_{xx} Y \sqrt{\frac{g_{tt}}{g_{rr}}}a'_z
+B q_2 h_{tz}\nonumber\\&&-\omega q_2 a_z=0. \label{connect}\nonumber
\eea
In order to deduce the RG flow equation of the transport, we define a matrix $\tilde{\sigma}$ by assuming
\[
\left \llbracket \begin{array}{c}
\sqrt{g_{tt}/g_{rr}} Z a_{z}^{\prime}\\
g^2_{xx}/\sqrt{g_{rr}g_{tt}}h_{tz}^{\prime}\\
\sqrt{g_{tt}/g_{rr}} Z g_{xx}{\chi}_z^{\prime}
\end{array}\right \rrbracket=\tilde{\sigma}\left \llbracket \begin{array}{c}
i\omega a_{z}\\
i\omega h_{tz}\\
i\omega {\chi}_z
\end{array}\right \rrbracket,
\]
where we have introduced the notation
\bea
\left \llbracket \begin{array}{c}
 a_{z}\\
 h_{tz}\\
 {\chi}_z
\end{array}\right \rrbracket\equiv
 \left(\begin{array}{ccc}
 a_{z} & a^{(2)}_{z} & a^{(3)}_{z}\\
h_{tz} & h^{(2)}_{tz} & h^{(3)}_{tz}\\
 {\chi}_{z} &{\chi}^{(2)}_{z} & {\chi}^{(3)}_{z}
\end{array}\right),
\eea
with $a^{(i)}_{z}$, $h^{(i)}_{tz}$ and ${\chi}^{(i)}_{z}$ linearly independent source vectors obeying the same equation as (\ref{connect}).  These auxiliary vectors are only introduced to make the source term to be a square matrix. So the notation $\llbracket ...\rrbracket$ is invertible in the following derivation, while the column matrix on the left hand side should also be understood in the same sense.
Taking derivative of $\tilde{\sigma}$  and using the EoM  repetitiously,  we obtain
\begin{eqnarray*}
\tilde{\sigma}^{\prime} & = & \left \llbracket\begin{array}{c}
\sqrt{g_{tt}/g_{rr}} Z a_{z}^{\prime}\\
g^2_{xx}/\sqrt{g_{rr}g_{tt}}h_{tz}^{\prime}\\
\sqrt{g_{tt}/g_{rr}} Z g_{xx}{\chi}_z^{\prime}
\end{array}\right \rrbracket \left \llbracket\begin{array}{c}
i\omega a_{z}\\
i\omega h_{tz}\\
i\omega {\chi}_z
\end{array}\right \rrbracket ^{-1}\nonumber\\&-&i\omega\tilde{\sigma}\left \llbracket\begin{array}{c}
 a_{z}^{\prime}\\
 h_{tz}^{\prime}\\
{\chi}_z^{\prime}
\end{array}\right \rrbracket \left \llbracket\begin{array}{c}
i\omega a_{z}\\
i\omega h_{tz}\\
i\omega {\chi}_z
\end{array}\right \rrbracket ^{-1}\\
 & = & \frac{\sqrt{g_{rr}/g_{tt}}}{i\omega}\left(\begin{array}{ccc}
-\omega^2 Z& \omega B Z&0\\
-\omega B Z & g_{xx}Y\beta^2+B^2 Z& i \omega g_{xx}Y\beta^2\\
0 & i \omega g_{xx}Y\beta^2& -\omega^2 g_{xx} Y
\end{array}\right)\nonumber\\&-&
i\omega \sqrt{\frac{g_{rr}}{g_{tt}}}\tilde{\sigma}\left(\begin{array}{ccc}
\frac{1}{Z}& 0 & 0\\
0 & \frac{g_{tt}}{g^2_{xx}}&0\\
0 & 0 & \frac{1}{g_{xx}Z}
\end{array}\right)\tilde{\sigma}\nonumber\\&+&\left(\begin{array}{ccc}
0& \frac{\sqrt{g_{rr}g_{tt}}}{g_{xx}} & 0\\
\frac{q_2}{Z}\sqrt{\frac{g_{rr}}{g_{tt}} }& 0&0\\
0 & 0 & 0
\end{array}\right)\tilde{\sigma}.
\end{eqnarray*}
The prime denotes the derivative with respect to $r$. At the event horizon, $g_{tt}/g_{rr}\rightarrow 0$ and $g_{tt} \sigma'/g_{rr} \rightarrow 0$ the regularity condition then requires
\begin{eqnarray}\label{sig0}
 \tilde{\sigma}_0=\left(\begin{array}{ccc}
Z & -\frac{ZB}{\omega}&0\\
\frac{BZ-iq_2}{\omega} & \sigma_{22} &-\frac{i  \beta g_{xx} Y}{\omega}\\
-\frac{i \beta g_{xx}Y}{\omega} & 0& g_{xx}Y
\end{array}\right).
\end{eqnarray}
Note that $\sigma_{22}$ can be determined by the constraint equation (\ref{connect}), which can be written as
\[
 \left(\begin{array}{ccc}
0 & 0&0\\
-iB & i \omega &-\beta\\
0 & 0& 0
\end{array}\right)\tilde{\sigma}=\left(\begin{array}{ccc}
0 & 0&0\\
q_2 & -Bq_2/\omega &0\\
0 & 0& 0
\end{array}\right).
\]
The above equation leads to
\bea
&&-i B \tilde{\sigma}_{11}+i\omega \tilde{\sigma}_{21}-\beta \tilde{\sigma}_{31}=q_2,\label{sig21}\\
&&-i B \tilde{\sigma}_{12}+i \omega \tilde{\sigma}_{22}-\beta \tilde{\sigma}_{32}=-\frac{B q_2}{\omega},\label{sig22}\\
&&-i B \tilde{\sigma}_{13}+i \omega \tilde{\sigma}_{23}-\beta \tilde{\sigma}_{33}=0. \label{sig23}
\eea
From (\ref{sig22}), we obtain
\bea
\tilde{\sigma}_{22}=\frac{i B q_2}{\omega^2}-\frac{Z B^2+g_{xx}Y \beta^2}{\omega^2}.
\eea
Equations (\ref{sig21}) and (\ref{sig23}) are evidently satisfied by components of $\tilde{\sigma}_0$ given in (\ref{sig0}).
Therefore, from the definition of $\tilde{\sigma}$, we have the regularity condition at event horizon
\be
\sqrt{\frac{g_{tt}}{g_{rr}}}a'_z\rightarrow i \omega a_z-i B h_{tz}\big|_{r=\rh}.
\ee
Together with the regularity condition for $h_{tz}$ from the first equation of (\ref{connect}), we arrive at
\bea
\!(\beta^2 g_{xx}Y+B^2 Z-i B q_2)h_{tz}\big|_{r=\rh}&=&(\omega B Z-i \omega q_2 ) a_z\big|_{r=\rh}\nonumber\\&-&i\omega g_{xx}Y \chi_z\big|_{r=\rh}.\!\nonumber
\eea
Keeping in mind that $h_{tz}=(h_{tx}-i h_{ty})/2$ and $\chi_z=(\chi_{x}-i \chi_{y})/2$, we easily obtain
\bea
h_{tx}&=&-i\omega\frac{a_x g_{xx}q_2 \beta^2 Y+a_y B(q_2^2+B^2 Z^2+g_{xx}Z \beta^2 Y)}{(B^2 Z+g_{xx} \beta^2 Y)^2+B^2 q_2^2}\nonumber\\
&-&\frac{Bg_{xx}Y\chi_y+(B^2 g_{xx}YZ+g^2_{xx}Y\beta^2)\chi_x}{(B^2 Z+g_{xx} \beta^2 Y)^2+B^2 q_2^2}\bigg|_{r=\rh},\nonumber\\
h_{ty}&=&i\omega\frac{a_x B(q_2^2+B^2 Z^2+g_{xx}Z \beta^2 Y)-a_y g_{xx}q_2 \beta^2 Y}{(B^2 Z+g_{xx} \beta^2 Y)^2+B^2 q_2^2}\nonumber\\
&+&\frac{Bg_{xx}Y\chi_x-(B^2 g_{xx}YZ+g^2_{xx}Y\beta^2)\chi_y}{(B^2 Z+g_{xx} \beta^2 Y)^2+B^2 q_2^2}\bigg|_{r=\rh},\nonumber
\eea
One can simply drop out the $\chi_i$ terms in the expressions for $h_{tx}$ and $h_{ty}$ since these terms do not contribute to the transport coefficients as shown below.
The radially conserved currents can be deduced from the Maxwell equation
\bea
&&J_{x}=i\omega Z a_x-q_2 h_{tx}-Z B h_{ty},\nonumber\\
&&J_{y}=i\omega Z a_y-q_2 h_{ty}+Z B h_{tx}.
\eea
The dc electric conductivity can be calculated via $\sigma_{ij}=\frac{\partial J_i}{\partial E_j}$, where $E_j=-i\omega a_j$.
The electrical conductivity tensor then
 reads as
\bea
&& \sigma_{xx}=\sigma_{yy}=\frac{\beta^2\rh^2(B^2+q_2^2 \rh^2+\beta^2 \rh^3)}{B^4+\beta^4 \rh^6+B^2 \rh^2(q_2^2+2\beta^2\rh)},\nonumber\\
&&\sigma_{xy}=-\sigma_{yx}=\frac{B q_2(B^2+q_2^2 \rh^2+2\beta^2 \rh^3)}{B^4+\beta^4 \rh^6+B^2 \rh^2(q_2^2+2\beta^2\rh)}.\nonumber
\eea
The radially conserved heat current are given by
\bea
&&\mathcal{Q}^x=-\sqrt{\frac{g_{tt}}{g_{rr}}}\frac{g'_{tt}}{g_{tt}}g_{xx}\delta h_{tx}\bigg|_{r=\rh},\nonumber\\
&&\mathcal{Q}^y=-\sqrt{\frac{g_{tt}}{g_{rr}}}\frac{g'_{tt}}{g_{tt}}g_{xx}\delta h_{ty}\bigg|_{r=\rh}.
\eea
So that we can evaluate the electrothermal conductivity matrix as
\bea \label{thermoelectric}
&&\alpha_{xx}=\alpha_{yy}=\frac{\partial \mathcal{Q}^x}{T \partial E_x}=\frac{4\pi q_2 \beta^2\rh^5}{B^2 q_2^2\rh^2+(B^2+\beta^2 \rh^3)^2},\\
&&\alpha_{xy}=-\alpha_{yx}=\frac{\partial \mathcal{Q}^y}{T \partial E_x}=\frac{4\pi B \rh(q_2^2\rh^2+B^2+\rh^3 \beta^2)}{B^2 q_2^2\rh^2+(B^2+\beta^2 \rh^3)^2}.\nonumber
\eea
In general, we have the relation
\begin{equation}
\label{pheno}
\left(\begin{array}{c} J_{i}  \\  Q_{i}  \end{array}\right)
=
\left(\begin{array}{cc}   {\sigma}_{ij} &{\alpha}_{ij} T \\  { \bar{\alpha}}_{ij} T & {\bar{\kappa}}_{ij} T \end{array}\right)
\left(\begin{array}{c} E_{j} \\ - (\nabla_{j} T)/T\end{array}\right)~.
\end{equation}
After obtaining the expressions for ${\sigma}_{ij}$,  ${\alpha}_{ij}$ and ${ \bar{\alpha}}_{ij}$, the thermal conductivity can be read off from (\ref{pheno}).
Considered the condition $Q_x=0, \nabla_{y} T=0$ and $E_y=0$, the thermal conductivity matrix component $\bar{\kappa}_{xx}$ is given by
\bea
\bar{\kappa}_{xx}=\frac{T \alpha^2_{xx}}{\sigma_{xx}-\sigma^0_{xx}},
\eea
where $\sigma^0_{xx}=\sigma_{xx}(q_2=0)$ denotes the quantum critical conductivity.
Similarly, the case $Q_x=0, \nabla_{x} T=0$ and $E_y=0$ gives
\be
\bar{\kappa}_{xy}=\frac{T \alpha_{xx}\alpha_{xy}}{\sigma_{xx}}.
\ee
Therefore, we finally obtain
\bea
&&\bar{\kappa}_{xx}=\frac{16\pi^2 T \rh^3(B^2+\beta^2 \rh^3)}{B^2 q_2^2\rh^2+(B^2+\beta^2 \rh^3)^2},\\
&&\bar{\kappa}_{xy}=\frac{16\pi^2 T B q_2\rh^4}{B^2 q_2^2\rh^2+(B^2+\beta^2 \rh^3)^2}.
\eea
These results  are consistent with the results obtained in \cite{donos,blake2015}.


\begin{thebibliography}{99}
\bibitem{convention} In conventional metals,  the Nernst signal is linear in the magnetic field.
\end{thebibliography}
\end{document}